\documentclass[11pt]{article}
\usepackage{amsmath}
\usepackage{graphicx}
\usepackage{color}
\usepackage{orcidlink}
\usepackage{caption}
\usepackage{subcaption}
\usepackage{hyperref}%\usepackage[colorlinks=true,linkcolor=blue]{hyperref}
\hypersetup{colorlinks=true, linkcolor=blue, citecolor=blue, urlcolor=blue}
\begin{document}
\baselineskip=16pt

\begin{center}
    {\large {\bf Impact of global monopole on heavy mesons in hot-dense medium }}
\end{center}

\vspace{0.2cm}

\begin{center}
    {\bf M. Abu-Shady\orcidlink{0000-0001-7077-7884}}\footnote{\bf dr.abushady@gmail.com}\\
    \vspace{0.1cm}
    {\it Department of Mathematics and Computer Science, Faculty of Science, Menoufia
University, Shbien El-Kom, Egypt}\\
    \vspace{0.3cm}
    {\bf Faizuddin Ahmed\orcidlink{0000-0003-2196-9622}}\footnote{\bf faizuddinahmed15@gmail.com (Corresponding author)}\\
    \vspace{0.1cm}
    {\it Department of Physics, University of Science \& Technology Meghalaya,
Ri-Bhoi, 793101, India}
\end{center}

\vspace{0.5cm}

\begin{abstract}
This research study is primarily focus on investigating how the topological effects influence the eigenvalue solutions in the presence of a hot-dense medium. To accomplish this, we employ the non-relativistic Schrödinger wave equation, taking into consideration both the quantum flux field and an interaction potential. Through this approach, we determine the energy eigenvalues and their corresponding wave functions using the Nikiforov-Uvarov method. Our findings indicate that when we consider both the topological effects and the magnetic flux, $\Phi$, there is a noticeable reduction in the binding energy within the hot-dense medium. Additionally, we analyze the role of the baryonic potential in shaping the binding energy within the $(T, u_b)$ plane. Interestingly, it is evident that the influence of the baryonic potential becomes more pronounced as its values decrease.
\end{abstract}

\vspace{0.1cm}


\section{Introduction}

The investigation of strongly interacting matter in extreme conditions has become a subject of great interest due to its relevance to particle physics and astrophysics. One specific area of importance is studying how the properties of hadrons, such as their masses, magnetic moments, and decay constants, can be altered when they propagate through a hot medium. Understanding the behavior of quarks and gluons in this hot medium, known as quark-gluon plasma (QGP), requires a thorough examination of hadron properties at finite temperature and density. The exploration of this phase, the QGP, is being conducted in experiments at RHIC (Brookhaven National Laboratory) and CERN, where there is substantial evidence supporting its existence \cite{ref1}. Numerous studies have been conducted on this topic, employing both relativistic and non-relativistic quark models as detailed in references \cite{ref2,ref3,ref4,ref5,ref6,ref7,ref8,ref9,ref10,ref11,ref12}. 

In Ref. \cite{ref2}, authors have been achieved an analytical solution for the $N$-radial Schrödinger equation by extending the Cornell potential to finite temperatures. The primary objective of this study was to investigate how the masses of charmonium and bottomonium states change when subjected to varying temperature conditions. In Ref. \cite{ref3}, authors focused on examining the dissociation of quarkonia, which are bound states formed by quarks and antiquarks, within a thermal quantum chromodynamics medium. They employed the conformable fractional version of the Nikiforov–Uvarov (CF-NU) method. This study aimed to shed light on how the thermal environment impacts the stability and existence of quarkonia states.

In Ref. \cite{ref4}, the primary focus centers on examining the thermodynamic properties of heavy mesons. These properties are determined by solving the $N$-dimensional radial Schr\"{o}dinger equation. This study calculated various thermodynamic characteristics of heavy mesons, shedding light on their behavior under different conditions. In Ref. \cite{ref5}, an analytical investigation of the N-radial Schrödinger equation was conducted, employing the methods of supersymmetric quantum mechanics. The study introduced the concept of a heavy-quarkonia potential at finite temperature and baryon chemical potential to assess its impact on the system's properties. Furthermore, Refs. \cite{ref6,ref7,ref8} collectively developed into the dissociation of quarkonia within an anisotropic plasma found in hot and dense media. These studies aim to comprehend how the anisotropic nature of the plasma affects the behavior and stability of quarkonia states.

The quark sigma model, a relativistic quark model, has emerged as a valuable tool in comprehending strong nuclear interactions \cite{ref9,ref10}. Within this model, the phenomenon of spontaneous chiral symmetry breaking and its restoration at higher temperatures are demonstrated. Numerous researchers have explored the Hartree approximation of the linear sigma model employing two or four quark flavors, investigating its behavior at different temperature regimes  \cite{ref1,ref11,ref12,ref13,ref14,ref15}. Furthermore, several studies have successfully applied the quark sigma model to characterize both static and dynamic baryons at various temperatures and densities, as documented in references \cite{ref16,ref17,ref18}. This demonstrates the model's versatility in describing the properties of baryons under diverse thermodynamic conditions.

Topological defects are exotic objects believed to have originated during phase transitions in the early universe \cite{AV,AV2}. They represent intriguing anomalies that emerge in a wide range of physical systems and have been extensively investigated in the realms of gravitation and cosmology, quantum mechanical system, atomic and molecular physics, and condensed matter physics. The literature recognizes various types of topological defects, among which cosmic strings \cite{AV,AV2,AV3} and global monopoles \cite{MB} have gained significant interest in theoretical physics. Each of these defects possesses distinctive properties and attributes, contributing significantly to our comprehension of fundamental physics. Cosmic strings, for instance, are stable one-dimensional defects theorized to have developed during a phase transition in the universe's early history. These strings are believed to play a pivotal role in shaping the large-scale structure of the cosmos by influencing the distribution of matter and the formation of galaxies.

Global monopoles \cite{MB} are intriguing three-dimensional spherical entities arising from the phenomenon of spontaneous symmetry breaking within specific grand unified theories. It possess unique characteristics, particularly concerning their mass and interactions, and they wield the potential to impact the cosmic evolution on cosmological scales. The physics of global monopole were discussed in Ref. \cite{BJP}. The exploration of topological defects has yielded profound insights into various facts of physics, fostering valuable connections between diverse fields of study. These defects, with their presence and behavior, hold significant implications not only for the early universe but also for understanding condensed matter systems and the interactions occurring at atomic and molecular levels. Several authors have been studied the quantum mechanical problems within the context of topological defects produced by cosmic strings and global monopoles (see, for references \cite{gg1, gg2, gg3, gg4, gg5, gg6, gg7, gg8}). Moreover, various potential models were considered in these investigations and obtained the eigenvalue solutions of the quantum system using different techniques or special functions. one of us has investigated the quantum dynamics of particles through the Schr\"{o}digner equation under the influence of a flux field in the presence of potential, such as pesudoharmonic, Kratzer, Yukawa, Deng-Fan, inverse quadratic Yukawa etc. in topological defect and obtained the eigenvalue solutions using the Nikiforov-Uvarov method (see, Refs. \cite{ref20,ref21,ref22,ref23,ref24,ref25,ref26,ref27,ref28}).

The primary goal of this research is to explore the effects of topological defect and magnetic flux field within a high density and temperature medium on heavy mesons. To accomplish this, we employ the Nikiforov-Uvarov (NU) method \cite{ref19} to solve the radial Schrödinger equation, enabling us to derive the energy eigenvalues and their corresponding wave functions. This NU method has widely been employed in solving quantum mechanical problems both in the relativistic and non-relativistic limit in the presence of different physical potentials by numerous researchers in the literature (see, for examples, \cite{hh1,hh2,hh3,hh4,hh5,hh6,hh7,hh8,hh9, rs1, rs2, rs3, rs4, rs5}). It's worth noting that, to the best of our knowledge, previous research has not sufficiently examined the impact of topological effects on heavy mesons within a hot-dense medium. Consequently, this study represents a significant and novel contribution to the field of research.

\section{Theoretical Description of the Nikiforov-Uvarov (NU) Method}

In this section, we will present a concise overview of the NU method \cite{ref19}, which serves as a valuable tool for solving second-order differential equations in the specified form given by
\begin{equation}
\Psi^{\prime\prime}(s)+\frac{\bar{\tau}(s)}{\sigma(s)}\Psi^{\prime}%
(s)+\frac{\tilde{\sigma}(s)}{\sigma^{2}(s)}\Psi(s)=0, \label{1}%
\end{equation}
where $\sigma(s)$ and $\tilde{\sigma}(s)$ are polynomials of maximum second degree and $\bar{\tau}(s)$ is a polynomial of maximum first degree with an appropriate $s=s(r)$ coordinate transformation. To find particular solution of Eq. (1) by separation of variables, if one deals with the transformation%
\begin{equation}
\Psi(s)=\Phi(s)\chi(s), \label{2}%
\end{equation}
it reduces to an equation of hypergeometric type as follows%
\begin{equation}
\sigma(s)\chi^{\prime\prime}(s)+\tau(s)\chi^{\prime}(s)+\lambda\chi(s)=0,
\label{3}%
\end{equation}
where%
\begin{equation}
\sigma(s)=\pi(s)\frac{\Phi(s)}{\Phi^{\prime}(s)}, \label{4}%
\end{equation}%
\begin{equation}
\tau(s)=\bar{\tau}(s)+2\pi(s);~\ \ \tau^{\prime}(s)<0, \label{5}%
\end{equation}
and
\begin{equation}
\lambda=\lambda_{n}=-n\tau^{\prime}(s)-\frac{n(n-1)}{2}\sigma^{\prime\prime}(s),n=0,1,2,... \label{6}%
\end{equation}
$\chi(s)=\chi_{n}(s)$ which is a polynomial of $n$ degree which satisfies the
hypergeometric equation, taking the following form%
\begin{equation}
\chi_{n}(s)=\frac{B_{n}}{\rho_{n}}\frac{d^{n}}{ds^{n}}(\sigma^{\prime\prime}(s)\rho(s)), \label{7}%
\end{equation}
where $B_{n}$ is a normalization constant and $\rho(s)$ is a weight function which satisfies the following equation
\begin{equation}
\frac{d}{ds}\omega(s)=\frac{\tau(s)}{\sigma(s)}\omega(s);\ \ \ \omega(s)=\sigma(s)\rho(s), \label{8}%
\end{equation}%
\begin{equation}
\pi(s)=\frac{\sigma^{\prime}(s)-\bar{\tau}(s)}{2}\pm\sqrt{(\frac{\sigma^{\prime}(s)-\bar{\tau}(s)}{2})^{2}-\tilde{\sigma}(s)+K\sigma(s),}
\label{9}%
\end{equation}
and
\begin{equation}
\lambda=K+\pi^{\prime}(s), \label{10}%
\end{equation}
the $\pi(s)$ is a polynomial of first degree. The values of $K$ in the square-root of Eq. (9) is possible to calculate if the expressions under the square root are square of expressions. This is possible if its discriminate is zero. (for detail, see Ref. \cite{ref19}).

The structure of this paper is as follows: In section 3, we consider the Schr\"{o}dinger wave equation under the influence of flux field in the presence of an interaction potential. Moreover, this quantum system is considered within the framework of the topological defect produced by a point-like global monopole and then in a hot-dense medium. We derive the radial equation of the wave and solve it using the Nikiforov-Uvarov method. In section 4, a thorough discussion of the results is presented, interpreting and analyzing the implications of the topological effects and the magnetic flux within a hot-dense medium on heavy mesons. Finally, in section 5,  we present our conclusions.

\section{Non-relativistic equation with potential under the effects of global monopole }

In this section, we aim to analysis the eigenvalues solution for non-relativistic quantum particles under the influence of a quantum flux field, taking into account the presence of a point-like global monopole under a potential in a hot-dense medium. Studies of the Schr\"{o}dinger wave equation in the background of a point-like global monopole with potential models have attracted much attention among researchers stated in the introduction. 

The radial Schr\"{o}dinger equation in the presence of a potential $V(r)$ under the influence of a flux field $\Phi_{AB}$ within the background of point-like global monopole is given by (see, Refs. \cite{ref20,ref21,ref22,ref23,ref24,ref25,ref26,ref27,ref28})
\begin{equation}
\frac{d^2\,\Psi(r)}{dr^2}+\Bigg[\frac{2\,M}{\alpha^2}\,\Big(E-V(r)\Big)-\frac{\iota^2}{r^2}\Bigg]\,\Psi(r)=0,\quad \iota^2=\frac{\ell^{\prime}(\ell^{\prime}+1)}{\alpha^2},
\label{11}
\end{equation}
where $\ell^{\prime}=(|m'|+\kappa)=(|m-\Phi|+\kappa)$ with $\kappa=0,1,2,3,...$, $m$ is the magnetic quantum number, $\ell$ is the angular momentum quantum number, $0 < \alpha < 1$ characterise the topological defect of point-like global monopole, and $\Phi=\frac{\Phi_{AB}}{\Phi_0}$ with $\Phi_0=2\,\pi\,e^{-1}$ is the amount of magnetic flux which is a positive real number.

In case of two-particle system interacting through a spherically symmetric potential $V(r)$, we replace mass $M$ of the quantum particles by reduced mass $\mu$ of the two-particle system, that is, $M \to \mu$. Therefore, the radial wave equation under the influence of the quantum flux field in this case can be written as 
\begin{equation}
\frac{d^2\,\Psi(r)}{dr^2}+\Bigg[\frac{2\,\mu}{\alpha^2}\,\Big(E-V(r)\Big)-\frac{\iota^2}{r^2}\Bigg]\,\Psi(r)=0,
\label{12}
\end{equation}
$\mu$ is the reduced mass for the quarkonium particle (for charmonium $\mu=\frac{m_{c}}{2}$ and for bottomonium $\mu=\frac{m_{b}}{2}$, respectively).

At finite temperature, interaction potential is given by \cite{ref30}
\begin{equation}
V(r)=a\,\left(m_{D},r\right) \,r-\frac{b\left(m_{D},r\right)}{r},
\label{13}
\end{equation}
where 
\begin{equation}
    a\left(T,r\right)=\frac{a}{m_{D}(T)\,r}\,\Big(1-e^{-m_{D}(T)\,r}\Big),\quad b\left(T, r\right)=b\,e^{-m_{D}(T)\,r}.
    \label{14}
\end{equation}
Here $m_{D}\left(T,u_{q}\right)$ is the Debye mass that vanishes at $T \to 0$, and $a, b$ are arbitrary constants that will be determined later (for detail, see Ref. \cite{ref30}). 

Thereby, substituting Eq. (\ref{13})--(\ref{14}) into the Eq. (\ref{12}) and using approximation $e^{-m_{D}\left(T\right)  r}=%
%TCIMACRO{\dsum \limits_{j=0}^{\infty}}%
%BeginExpansion
{\displaystyle\sum\limits_{j=0}^{\infty}}
%EndExpansion
\frac{\left(  -m_{D}\left(  T\right)  r\right)  ^{j}}{j!}$ up to second-order, which gives a good accuracy when $m_{D}r\ll1$, we obtain%
\begin{equation}
\Bigg[\frac{d^{2}}{dr^{2}}+2\mu^{\prime}\Bigg(E-A+\frac{b}{r}-Cr+Dr^{2}-\frac{\ell^{\prime}(\ell^{\prime}+1)}{2\,\mu'\,r^{2}}\Bigg)\Bigg]\,R(r)=0,\label{15}%
\end{equation}
where,
\begin{equation}
A=b~m_{D}\left(  T\right),\quad C=a-\frac{1}{2}\,b\,m_{D}^{2}\left(T\right),\quad D=\frac{1}{2}a~m_{D}\left(  T\right),\quad    \mu^{\prime}=\frac{\mu
}{\alpha}.\label{16}
\end{equation}

By taking $r=\frac{1}{x}$, Eq. (\ref{15}) takes the following form%
\begin{equation}
\Bigg[\frac{d^{2}}{dx^{2}}+\frac{2}{x}\frac{d}{dx}+\frac{2\,\mu^{\prime}}{x^{4}}\Big(E-A+bx-\frac{C}{x}+\frac{D}{x^{2}}-\frac{\ell^{\prime}(\ell^{\prime}
+1)}{2\mu}x\Big)\Bigg]\,R(x)=0. \label{17}%
\end{equation}
The scheme is based on the expansion of $\frac{C}{x}$ and $\frac{D}{x^{2}}$ in a power series around the characteristic radius $r_{0}$ of meson up to the second order. Setting $y=x-\delta$, where $\delta=\frac{1}{r_{0}}$, thus, we expand the $\frac{c}{x}$ and $\frac{D}{x^{2}}$ into a series of powers around $y=0$.%

\begin{align}
\frac{C}{x}=\frac{C}{y+\delta}=\frac{1}{\delta}\Big(1+\frac{y}{\delta}\Big)^{-1}=\frac{C}{\delta}\Big(1-\frac{y}{\delta}+\frac{y}{\delta^{2}}\Big)=C\Big(\frac{3}{\delta}-\frac{3x}{\delta^{2}}+\frac{x^{2}}{\delta^{3}}\Big)\text{.}
\label{18}
\end{align}

Similarly,%

\begin{equation}
\frac{D}{x^{2}}=D\Big(\frac{6}{\delta^{2}}-\frac{8x}{\delta^{3}}+\frac{3x^{2}%
}{\delta^{4}}\Big)\text{.} \label{19}%
\end{equation}

By substituting Eqs. (\ref{18})--(\ref{19}) into the Eq. (\ref{17}), we obtain the following second-order differential equation form given by
\begin{equation}
\Bigg[\frac{d^{2}}{dx^{2}}+\frac{2}{x}\frac{d}{dx}+\frac{2\mu^{\prime}%
}{x^{4}}(-A_{1}+A_{2}x-A_{3}x^{2})\Bigg]\,R(x)=0,\label{20}%
\end{equation}
where, 
\begin{equation}
A_{1}=-\Big(E-A-\frac{3C}{\delta}+\frac{6D}{\delta^{2}}\Big),\quad A_{2}=\Big(\frac{3C}{\delta^{2}}-\frac{8D}{\delta^{3}}+b\Big),\quad  A_{3}=\Big(\frac{C}{\delta^{3}}-\frac{3D}{\delta^{4}}+\frac{\ell'\,(\ell'+1)}{2\mu}\Big).\label{21}   
\end{equation}
%$\frac{1}{x}$ expansion gives a good accuracy when $\delta$ tends to $x$.

By comparing Eq. (\ref{20}) and Eq. (\ref{1}), we find $\bar{\tau}(s)=2x$, $\sigma
(s)=x^{2}$, and $\tilde{\sigma}(s)=2\mu^{\prime}(-A_{1}+A_{2}x-A_{3}x^{2})$. By following the NU method that mentioned in Sec. 2, therefore%
\begin{equation}
\pi=\pm\sqrt{\left(  K+2A_{3}\right)  x^{2}-2A_{2}x+2A_{1}}.\label{22}%
\end{equation}
The constant $K$ is chosen such as the function under the square root has a double zero, i.e. its discriminant $\Delta=4A_{2}^{2}-8A_{1}\left(
K+2A_{3}\right)  =0$. Hence,
\begin{equation}
\pi=\pm\frac{1}{\sqrt{2A_{1}}}\left(  2A_{1}-A_{2}x\right)  \text{.}\label{23}%
\end{equation}
Thus,
\begin{equation}
\tau=2x\pm\frac{1}{\sqrt{2A_{1}}}\left(  2A_{1}-A_{2}x\right)  \text{.}%
\label{24}%
\end{equation}
For bound state solutions, we choose the positive sign in above equation so
that the derivative
\begin{equation}
\tau^{\prime}=2-\frac{2A_{2}}{\sqrt{2A_{1}}}\text{.}\label{25}%
\end{equation}
By using Eq. (\ref{10}), we obtain%
\begin{equation}
\lambda=\frac{A_{2}^{2}}{2A_{1}}-2A_{3}-\frac{A_{2}}{\sqrt{2A_{1}}},\label{26}%
\end{equation}
and Eq. (\ref{6}), we obtain
\begin{equation}
\lambda_{n}=-n\left(  2-\frac{2A_{2}}{\sqrt{2A_{1}}}\right)  -n(n-1)\text{.}%
\label{27}%
\end{equation}
From Eq. (\ref{6}); $\lambda=\lambda_{n}$. The energy eigenvalues of Eq. (\ref{15}) in the
hot-dense medium is given
\begin{equation}
E_{n\,\ell}^{N}=A+\frac{3C}{\delta}-\frac{6D}{\delta^{2}}-\frac{2\mu^{\prime}%
(\frac{3C}{\delta^{2}}+b-\frac{8~D}{\delta^{3}})^{2}}{\Big[(2n+1)\pm\sqrt
{1+\frac{8\mu^{\prime}C}{\delta^{3}}+\frac{4}{\alpha}\,\ell^{\prime}(\ell^{\prime
}+1)-\frac{24\mu^{\prime}D}{\delta^{4}}}\Big]^{2}}.\label{28}%
\end{equation}
The radial of wave function of Eq. (\ref{15}) takes the following form
\begin{equation}
R_{n\,\ell}\left(  r\right)  =C_{n\,\ell}~r^{-\frac{A_{2}}{\sqrt{2A_{1}}}-1}%
e^{\sqrt{2A_{1}}\,r}(-r^{2}\frac{d}{d\,r})^{n}(r^{-2n+\frac{A_{2}}{\sqrt{2A_{1}}}%
}e^{-2\sqrt{2A_{1}}r}).\label{29}%
\end{equation}
$C_{nL}$ is the normalization constant that is determined \ by $\int\left\vert
R_{nL}\left(  r\right)  \right\vert ^{2}dr=1$.

\begin{figure}
    \centering
    \includegraphics[width=0.65\textwidth,height=2.2in]{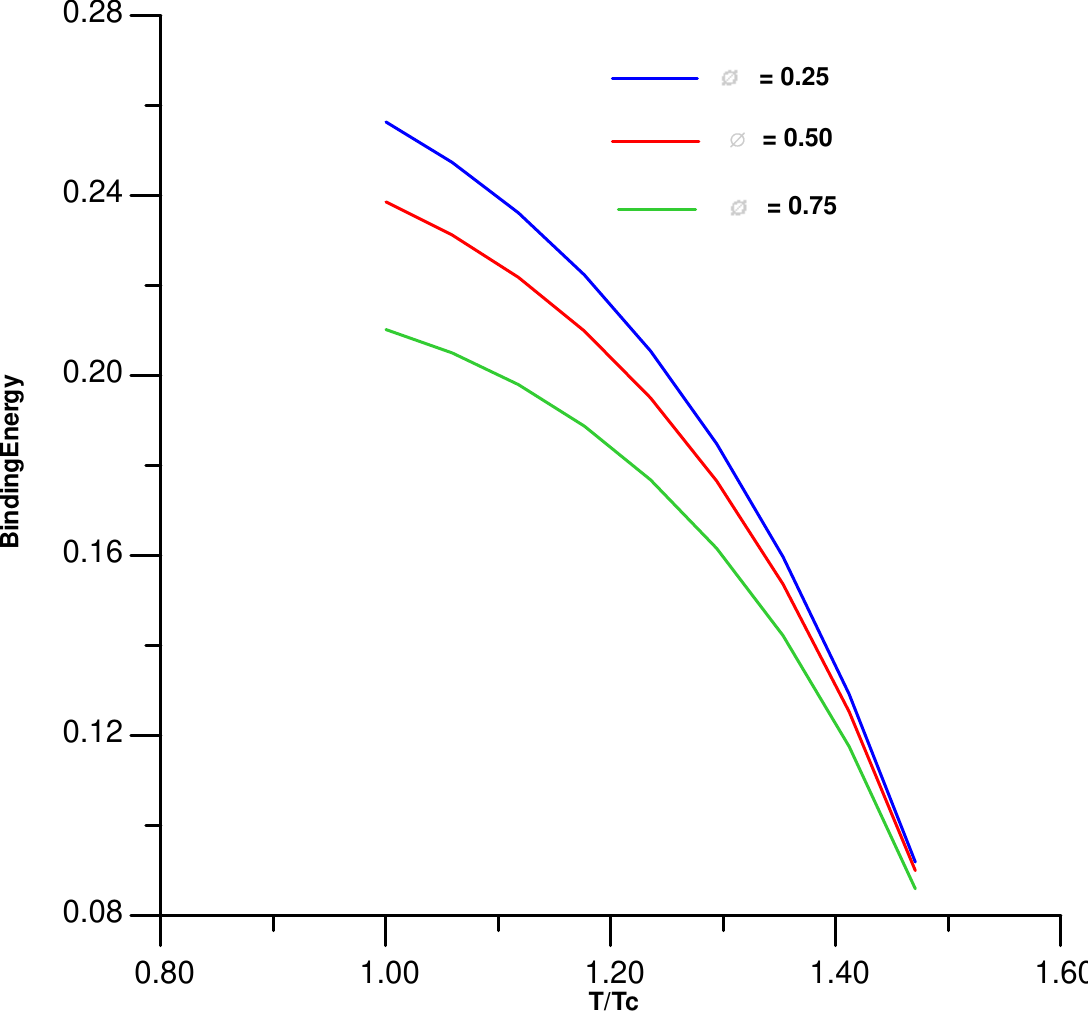}
    \caption{The binding energy is plotted as a function of $\frac{T}{T_{c}}$ for different values of $\Phi=0.25 (\rm{Blue}), 0.50 (\rm{Red}),0.75(\rm{Green})$ at u$_{b}=0$ and $\alpha=1.0$}
    \label{fig: 1}
    \hfill\\
    \includegraphics[width=0.65\textwidth,height=2.2in]{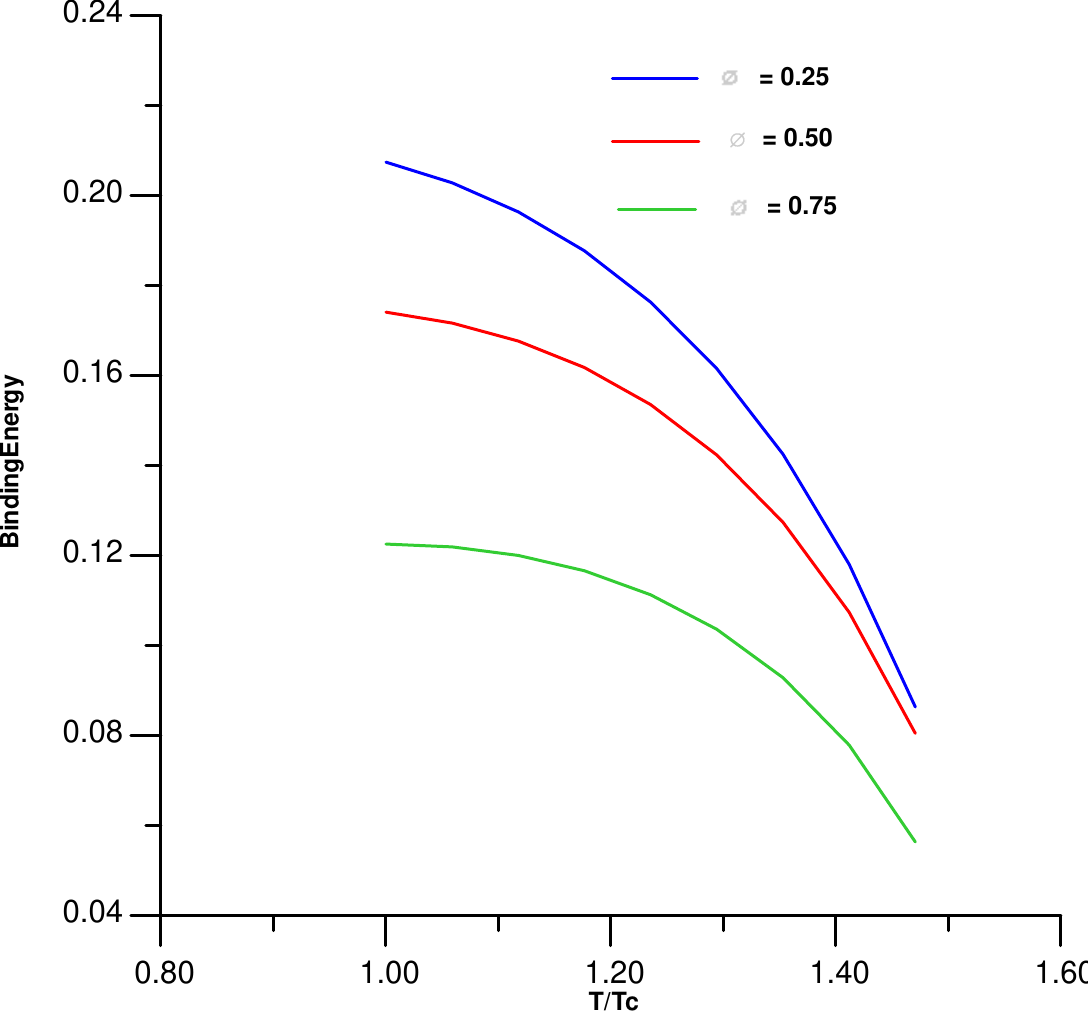}
    \caption{The binding energy is plotted as a function of $\frac{T}{T_{c}}$ for different values of $\Phi=0.25 (\rm{Blue}), 0.50 (\rm{Red}),0.75 (\rm{Green})$ at u$_{b}=0$ and $\alpha=0.25$}
    \label{fig: 2}
    \hfill\\
    \includegraphics[width=0.7\textwidth,height=2.2in]{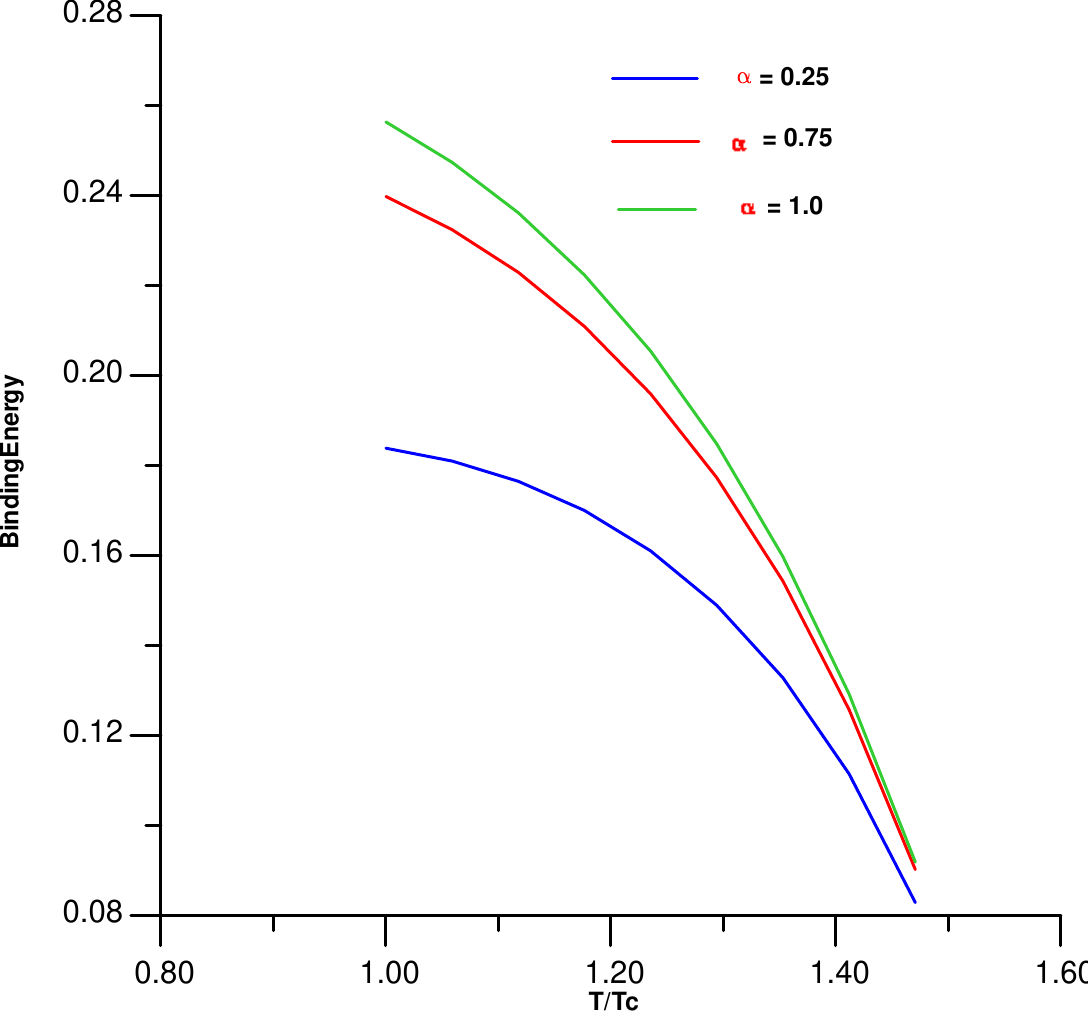}
    \caption{The binding energy is plotted as a function of $\frac{T}{T_{c}}$ for different values of $\alpha=0.25 (\rm{Blue}), 0.75 (\rm{Red}), 1.0 (\rm{Green})$ at u$_{b}=0$ and $\Phi=0.25$}
    \label{fig: 3}
\end{figure}

\begin{figure}
    \centering
    \includegraphics[width=0.65\textwidth,height=2.7in]{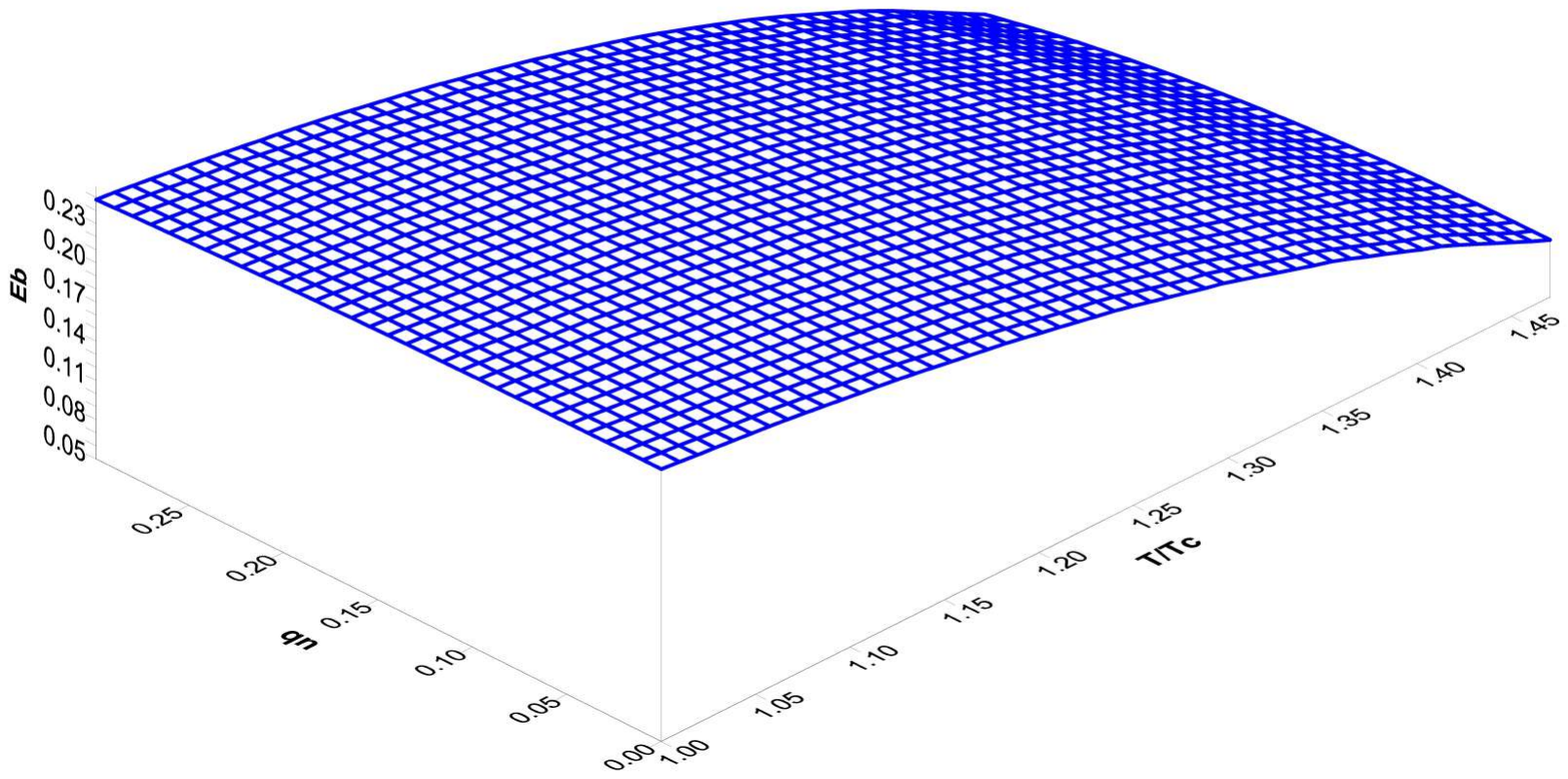}
    \caption{The binding energy is plotted as a function of u $_{b}$ and $\frac{T}{T_{c}}$ at $\Phi=0.25$ and $\alpha=1$}
    \label{fig: 4}
    \hfill\\
    \includegraphics[width=0.65\textwidth,height=2.7in]{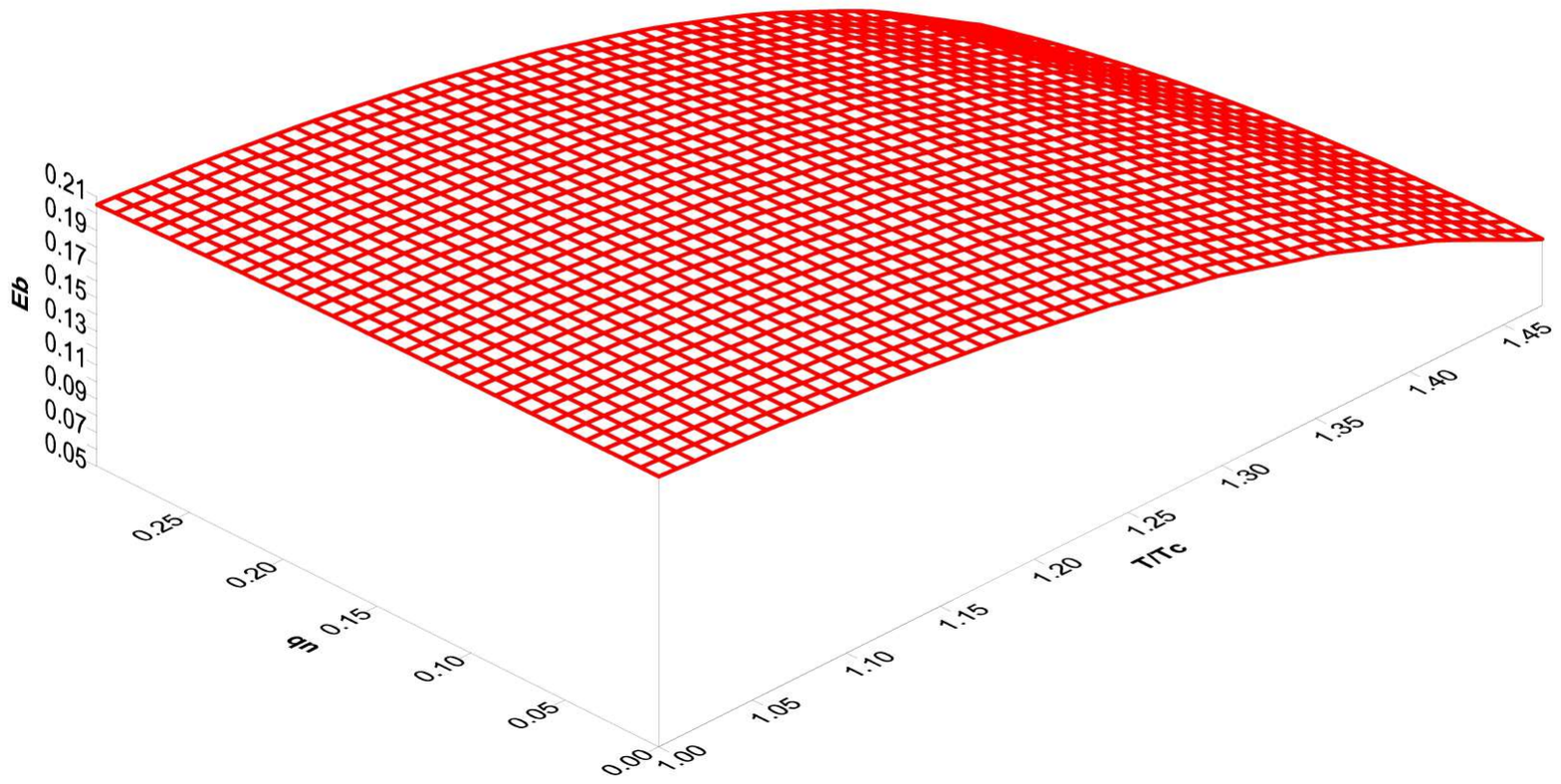}
    \caption{The binding energy is plotted as a function of u $_{b}$ and $\frac{T}{T_{c}}$ at $\Phi=0.25$ and $\alpha=0.4$}
    \label{fig: 5}
    \hfill\\
    \includegraphics[width=0.65\textwidth,height=2.7in]{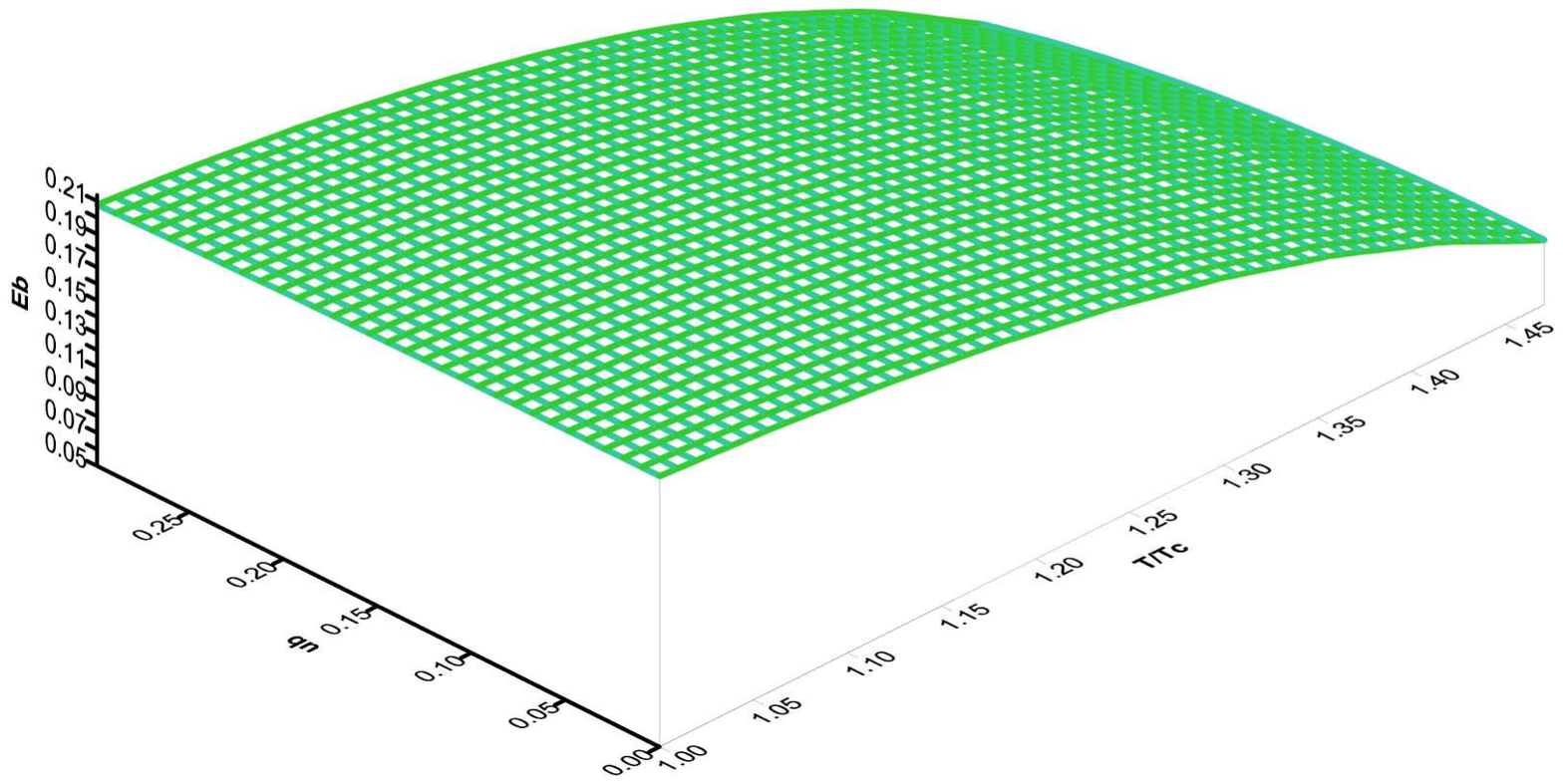}
    \caption{The binding energy is plotted as a function of u$_{b}$ and $\frac{T}{T_{c}}$ at $\Phi=0.75$ and $\alpha=1$}
    \label{fig: 6}
\end{figure}

\begin{figure}
    \centering
    \includegraphics[width=0.6\textwidth,height=2.8in]{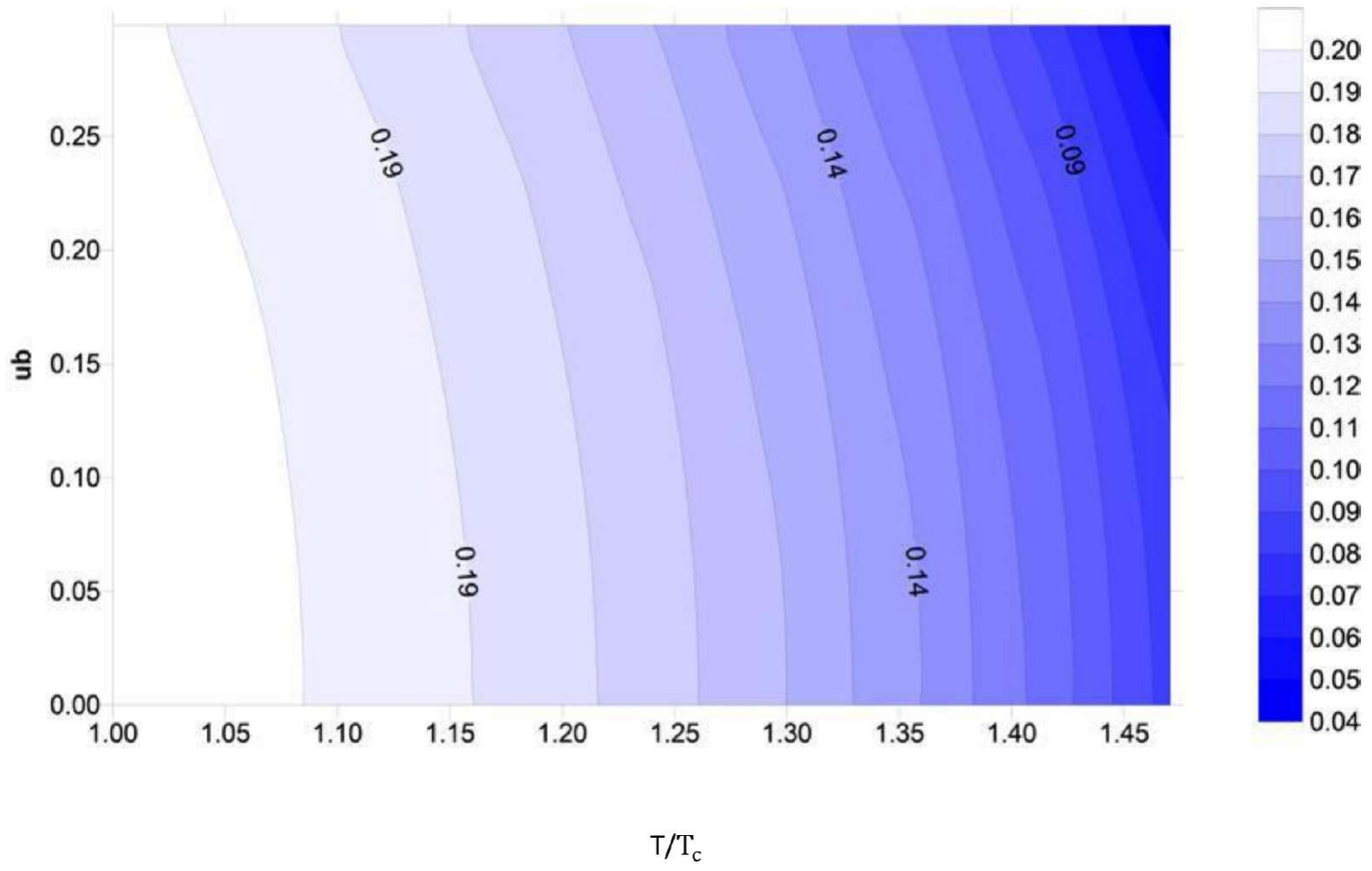}
    \caption{The binding energy is plotted as a function of u$_{b}$ and $\frac{T}{T_{c}}$ at $\Phi=0.25$ and $\alpha=0.4$}
    \label{fig: 7}
    \hfill\\
    \includegraphics[width=0.6\textwidth,height=2.8in]{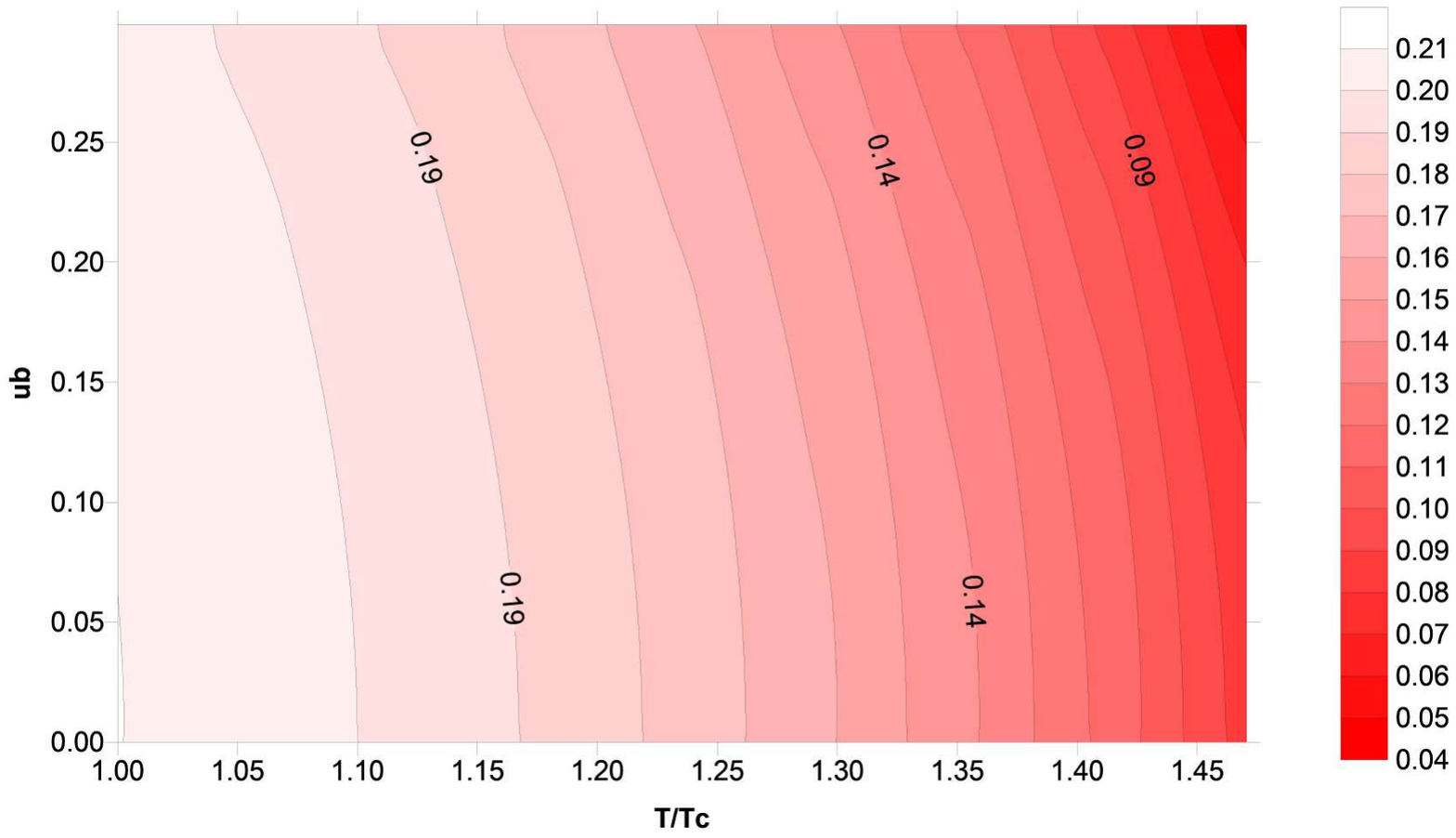}
    \caption{The binding energy is plotted as a function of u$_{b}$ and $\frac{T}{T_{c}}$ at $\Phi=0.75$ and $\alpha=1.0$}
    \label{fig: 8}
\end{figure}

\section{Discussion of Results}

In this section, we calculate spectra of the heavy quarkonium system such as bottomonium mesons in the hot and dense madium. The mass of quarkonium is calculated in the 3-dimensional space. We apply the following relation as in Ref. \cite{ref2}
\begin{equation}
M=2\,m+E_{n \ell}, \label{30}%
\end{equation}
where $m$ is quarkonium bare mass for the charmonium or bottomonium mesons. 

By using Eq. (\ref{28}), we write Eq. (\ref{30}) as follows:%
\begin{equation}
M=2m+A+\frac{3C}{\delta}-\frac{6D}{\delta^{2}}-\frac{2\mu^{\prime}(\frac
{3C}{\delta^{2}}+b-\frac{8~D}{\delta^{3}})^{2}}{\Big[(2n+1)\pm\sqrt{1+\frac
{8\mu^{\prime}C}{\delta^{3}}+\frac{4}{\alpha}\ell^{\prime}(\ell^{\prime}%
+1)-\frac{24\mu^{\prime}D}{\delta^{4}}}\Big]^{2}}\label{31}%
\end{equation}
Equation (\ref{31}) represents the quarkonium masses in hot and dense medium with topological effects and magnetic flux. By taking $\alpha=1$and $\Phi=0,$ we obtain 
\bigskip%
\begin{equation}
M=2m+A+\frac{3C}{\delta}-\frac{6D}{\delta^{2}}-\frac{2\mu(\frac{3C}{\delta
^{2}}+b-\frac{8~D}{\delta^{3}})^{2}}{\Big[(2n+1)\pm\sqrt{1+\frac{8\mu C}%
{\delta^{3}}+4\ell(\ell+1)-\frac{24\mu D}{\delta^{4}}}\Big]^{2}} \label{32}%
\end{equation}
\bigskip Eq. (\ref{32}) coincides with the result obtained in Ref. $\left[  2\right]$.

We can obtain the quarkonium masses at classical case by taking $T=0$ leads to $A=D=0$ and $C=a$, $\alpha=1$, and $\Phi=0$. Therefore, Eq. (\ref{31}) takes the following form%
\begin{equation}
M=2m+\frac{3a}{\delta}-\frac{2\mu(\frac{3a}{\delta^{2}}+b)^{2}}{\Big[(2n+1)\pm
\sqrt{1+\frac{8\mu a}{\delta^{3}}+4\ell(\ell+1)}\Big]^{2}}. \label{33}%
\end{equation}

Equation (\ref{33}) coincides with the result obtained in Ref. \cite{ref31}. In the present work, the Debye mass $D(T,\mu_{b})$ is given in Refs. \cite{ref32,ref33}%
\begin{equation}
D(T,\mu_{b})=gT\sqrt{\frac{N_{c}}{3}+\frac{N_{f}}{6}+\frac{N_{f}}{2\pi^{2}%
}\left(  \frac{\mu_{q}}{T}\right)  ^{2}}, \label{34}%
\end{equation}
where, $g$ \ is the coupling constant as defined in Ref. \cite{ref34}, $\mu_{q}$ is the quark chemical potential $\left(  \mu_{q}=\frac{\mu_{b}}{3}\right)  $, $N_{f}$ \ is number of flavours, and $N_{c}$ is number of colors.%

For the bottomonium meson, the binding energy is plotted as a ratio of temperature ($\frac{T}{T_{a}}$) where $T_{a}$ is a critical temperature in Figure 1. This plot is done when the baryonic chemical potential is not considered. It is observed that the binding energy decreases as the temperature increases. This behavior is similar for different values of the magnetic flux ($\Phi$). When the magnetic flux ($\Phi$) is decreased, the curves on the plot shift to higher values. Furthermore, the effect of temperature is more significant at the critical temperature ($T_{a}=0.17$ GeV). As the temperature increases beyond this critical temperature, the effect of magnetic flux ($\Phi$) becomes less pronounced in Figure 2. The authors of Refs. \cite{ref20,ref22,ref24,ref25,ref28} considered the topological effects without considering the hot-dense medium and found that the potential energy of the system is shifted to higher values with increasing values of magnetic flux ($\Phi$) and the global monopole parameter $\alpha$, which is consistent with our findings. Fig. 3 shows the binding energy plotted for different values of $\alpha$ without considering the baryonic chemical potential ($u_{b}=0$). It is noted that the binding energy reaches its maximum when the topological effects are ignored, that is,  $\alpha=1$. Thus, one can see that the values of the binding energy changes with the topological parameter $0 < \alpha<1$. The curves in Figure 3 are shifted from each other when topological effects are considered at $\alpha=0.25$. Overall, these findings suggest that the topological effects, magnetic flux ($\Phi$), and parameter $\alpha$ have significant impacts on the binding energy of the bottomonium meson in different temperature regimes, as illustrated above. In Figure 4, we plotted the binding energy as a function of ratio of temperature and the baryonic chemical potential, in which we study the effect of dense medium on the binding energy in the hot medium. When ignored the effect of topological effects at $\alpha=1$ with magnetic flux $\Phi=0.25,$ we note that the binding energy decreases with increasing the temperature at any value of the baryonic chemical potential. Additionally, the binding energy decreases slowly by increasing baryonic chemical potential. Therefore, we deduced that the effect of hot medium is more effect on the binding energy of bottomonium. In Figure 5, we consider the topological effects at $\alpha=0.25,$ we obtain a similar behavior as in Figure 4, but we note the binding energy decreases when the topological effects are considered. In Figure 6, by fixing $\alpha=1$ and increases $\Phi=0.75$, we note the binding energy is a little small in comparison with Figure 4. Figure $\left(  \text{7}\right)  $ show that the binding energy of bottomonium at vanishing of baryonic potential is greater than the binding energy at higher values of baryonic chemical potential in which the topological effects $\alpha=0.4$ and magnetic flux $\Phi=0.25$ are taken. By comparing with Figure $\left(  8\right)  $, we note that the binding energy increases by ignoring the the topological effects $\alpha=1.0$. Additionally, we note that the regular change in the binding energy in the plane of $\left(  T,u_{b}\right)  $ above the critical temperature $T_{c}$. In Figures 7-8, we have illustrated the binding energy with and without the effects of global monopole parameter $\alpha$ for a fixed value of magnetic quantum flux $\Phi$.

\section{Conclusions}

Our primary objective was to analyze the effects of topological phenomena induced by a point-like global monopole in the presence of a hot-dense medium. To achieve this, we solved the Schrödinger wave equation under the influence of a quantum flux fields, incorporating an interaction potential. Utilizing the parametric Nikiforov-Uvarov method, we successfully obtained the energy eigenvalues and corresponding wave functions of the particles.

The results revealed that both the topological defect parameter, denoted as $\alpha$, and the magnetic flux, represented by $\Phi$, had a significant influence on the eigenvalues when subjected to a hot-dense medium. This observation underscores the importance of considering these factors when studying such systems. Furthermore, we explored the impact of the baryonic potential on the binding energy in the $(T, u_b)$ plane. Interestingly, we observed that the effect of the baryonic potential was more pronounced when its values were smaller. This finding highlights the sensitivity of the system to changes in the baryonic potential and its potential implications for understanding the system's behavior. Overall, our study provides valuable insights into the intricate interplay of topological effects and magnetic flux within a hot-dense medium. By shedding light on these complex interactions, our findings contribute to a deeper understanding of the system's behavior and pave the way for further research in this fascinating area of study.

Our investigation involved a comprehensive theoretical examination of the influence of several critical factors, such as the topological defect produced by a point-like global monopole, and the magnetic flux, particularly in the context of a hot and/or dense medium. By considering these factors, we gained valuable insights into the behavior and properties of the quantum system under investigation, shedding light on the intricate interplay between the quantum flux field, the topological defect, and the binding energy of the quarkonium system.

\section*{Acknowledgements}

We would like to thank the anonymous referees' for their helpful suggestions and comments. FA acknowledges the Inter University Centre for Astronomy and Astrophysics (IUCAA), Pune, India for granting visiting associateship.

\section*{Data Availability}

No data were generated or analyzed in this study.

\section*{Conflicts of Interests}

Authors declares no such conflict of interests.

\end{document}